# THE MESA SECURITY MODEL 2.0: A DYNAMIC FRAMEWORK FOR MITIGATING STEALTH DATA EXFILTRATION

Sanjeev Pratap Singh and Naveed Afzal

Cybersecurity Analytics, Takeda Pharmaceuticals, Cambridge MA, USA

*ABSTRACT*

*The rising complexity of cyber threats calls for a comprehensive reassessment of current security frameworks in business environments. This research focuses on Stealth Data Exfiltration (SDE), a significant cyber threat characterized by covert infiltration, extended undetectability, and unauthorized dissemination of confidential data. Our findings reveal that conventional defense-in-depth strategies often fall short in combating these sophisticated threats, highlighting the immediate need for a shift in information risk management across businesses. The evolving nature of cyber threats, driven by advancements in techniques, such as social engineering, multi-vector attacks, and the emergence of Generative AI, underscores the need for robust, adaptable, and comprehensive security strategies. As we continue to navigate this complex landscape, it is crucial that we stay ahead of the curve, anticipating potential threats, and continually updating our defenses to protect against them.*

*We propose a shift from traditional perimeter-based, prevention-focused models, which depend on a static attack surface, to a more dynamic framework that prepares for inevitable breaches. This suggested model, known as 'MESA 2.0 Security Model', prioritizes swift detection, immediate response, and ongoing resilience, thereby enhancing an organization's ability to promptly identify and neutralize threats, significantly reducing the consequences of security breaches. This study suggests that businesses adopt a forward-thinking and adaptable approach to security management, which is crucial for staying ahead of the ever-changing cyber threat landscape. By shifting focus from merely preventing incidents to effectively managing them, organizations can better safeguard their vital digital assets against the increasingly complex tactics used by contemporary cyber adversaries. This study provides valuable insights and a solid strategic framework that aims to steer the development of future security practices and policies to effectively address and mitigate advanced persistent threats.*

*KEYWORDS*

*Stealth Data Exfiltration (SDE), MESA Security Model, Social Engineering, Multi-Vector Attacks, Advanced Persistent Threats (APTs), Zero-Day Vulnerabilities, Threat Intelligence, Cybersecurity Frameworks, Data Classification and Asset Inventory, Network Traffic Analysis*

## 1. INTRODUCTION

The cybersecurity landscape has seen significant changes in recent years. Today's malware and exploits are increasingly sophisticated and well-coordinated, challenging traditional security models and techniques. As defenses at the perimeter become less effective, it is important to reconsider and potentially transform the management of security, threats, and information risks in enterprises. The threats addressed in this study exhibit three key characteristics: using social engineering to infiltrate infrastructure, remaining hidden in networks for extended periods to monitor activities and gather data, and eventually transferring information to an external party.

Cybersecurity's evolving threat landscape poses significant challenges to information security practitioners, who now face advanced, well-funded threats from organizations backed by nation-states or criminal enterprises. These sophisticated attacks often target human vulnerabilities through social engineering and can remain undetected for extended periods while launching coordinated assaults across multiple channels. Employing tactics like masquerading as legitimate users and exploiting backdoors and zero-day vulnerabilities, these threats are exceedingly difficult to detect and mitigate [1].As malware continues to evolve in complexity, an approach is required to manage it. The rising sophistication of threats requires a shift from traditional security measures to more dynamic strategies that incorporate cutting-edge technologies and a thorough understanding of both the cyberthreat landscape and the specific tactics used by attackers. This includes the use of advanced persistent threat (APT) models that focus on complex, continuous, and targeted attacks [2].

To effectively tackle these challenges, organizations must adopt a multifaceted approach that includes updating their threat models to address high-level threats, implementing stronger detection systems, and fostering a deeper understanding of the tactics employed by modern adversaries. This holistic approach is essential for protecting valuable digital assets and ensuring the resilience of organizational infrastructure against today's sophisticated cyber threats. Organizations must confront the reality of sophisticated attacks breaching current defenses and recognize the need for new approaches to minimize risk and data loss. A model that assumes a breach is unavoidable could focus on proactive measures to reduce the impact of an attack. This study aims to develop a novel framework to enhance an organization's resilience to emerging security threats:

- Characterizing the threat model, researching recent threats and thorough analysis of a typical attack, including social engineering, technical reconnaissance, and data exfiltration components.
- Identifying gaps in current administrative, operational, and technical controls. Introducing additional protective mechanisms to supplement and strengthen existing security models.

This study identifies the strengths and shortcomings of current enterprise security approaches and introduces new methods and technologies that could bolster an enterprise's security measures against this constantly evolving threat [3]. The goal of this study is to establish procedures, technologies, and a framework that effectively counteracts this insidious threat and ultimately lays the groundwork for a paradigm shift in the perception and management of data security in enterprises.

## 2. ANATOMY OF A STEALTH DATA EXFILTRATION ATTACK

Although each Stealth Data Exfiltration (SDE) differs in the specific technology and threat vectors used, the studied exploits exhibit a common modus operandi. Refer to Figure 1 for further details.

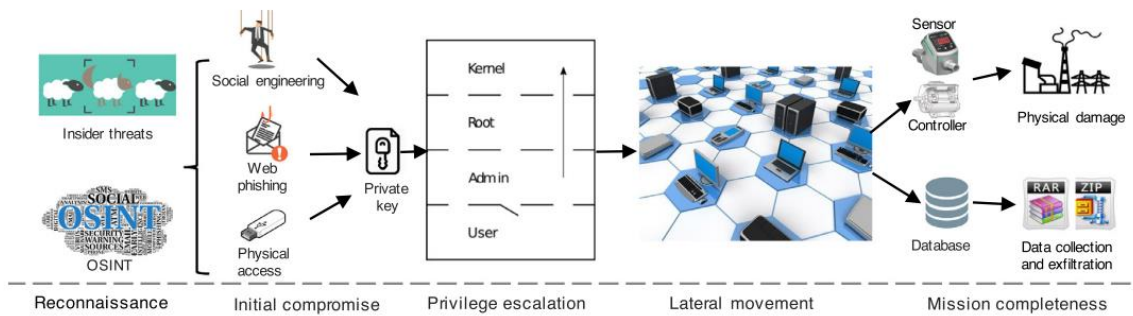

Figure 1 – Multi-stage structure of APTs includes stages like reconnaissance, initial compromise, privilege escalation, lateral movement, and mission execution of SDE[4]

It starts with reconnaissance that is targeted using social networks to devise social engineering schemes to acquire network access. Once access is obtained, the malware operates clandestinely and covertly within the network, keeping tabs on activities, enhancing privileges, investigating network resources, targeting digital assets, and pursuing other low-risk exploits. The final phase entails communication back to the command-and-control center to get additional instructions, and ultimately, transfer digital assets. SDE is a sophisticated attack that raises the stakes and significantly changes the threat landscape in terms of its impact on business, level of sophistication, and motives [5], [6].

## 2.1. Phase 1 - Social Engineering

Attackers exploit their inherent human predisposition to trust and assist others. This exploitation of human nature is a significant aspect of these attacks. Organizations typically allocate significant resources to implementing robust security measures. However, a single compromised individual can easily circumvent these defenses, rendering the defense-in-depth architecture ineffective. After all, humans remain our greatest vulnerability.

With the proliferation of social networks, social engineering has become exponentially easier than in the past. Troves of personal information are readily available through platforms such as Facebook, LinkedIn, Twitter, and thousands of affinity-group networks. These resources serve as starting points for social engineers. For instance, an attacker may identify an organization and conduct a search on LinkedIn for individuals who may have system administrative privileges within the organization. Once these individuals are identified, additional personal information about the intended targets can be obtained through Facebook and other websites[7], [8].Armed with knowledge of a person's occupation, friends, and interests, an attacker can craft a highly relevant and enticing email, known as a spear-phishing attack[9], [10]. These emails specifically target the victim and carry a malicious payload in the form of an attachment or link. Microsoft documents and PDFs are common attack vectors; however, attackers also conceal malware in images using steganography[11].

The objective of a spear-phishing attack is to lure the victim to open the attachment (or click on the link) and execute the initial malware. Once executed, the malicious payload installs malware on the compromised machine and establishes itself quietly on the network. Thus, the social engineering dimension of the SDE attack is successfully completed.

## 2.2. Phase 2 - Technical Reconnaissance

As previously stated, this study aims to establish a model that will assist enterprises in combatting this evolving threat. Although each SDE attack is unique and levels of 'stealthiness' vary, they all perform operations and leave traces of their activity. Studying and understanding these fingerprints form the basis for this new, improved security model. Once the initial malware is loaded and a covert presence is established, SDEs share some general traits and modes of operation that can be used to establish a framework for research and analysis.

By leveraging this simple framework to study electronic fingerprints, one can gain a greater understanding of an attacker and its malicious activities.

- **Escalate privileges**: The initial objective after gaining access to the network is to elevate privileges. The goal is to acquire administrative credentials. A stealthy approach is to target system administrators during the social engineering phase. If social engineering is unavailable, attackers may exploit configuration errors, low-level kernel vulnerabilities, and software bugs to escalate privileges. Evidence of such activities includes modifying access and authorization levels as well as changing password and privilege files without a corresponding ticket.
- **Harvesting Credentials**: In the pursuit of network privileges, attackers may deploy keystroke loggers to capture user IDs and passwords in real-time[12], [13]. Alternatively, they may transfer the access control data offsite to the Command and Control (CnC) center, where brute force analysis can be conducted undetected[14], [15].Due to the prevalence of data breaches, there is an abundance of stolen credentials that can be easily accessed by attackers. This has resulted in the rise of a new type of attack known as Account Takeover (ATO)[16]. In the past, attackers would employ offline bruteforce analysis to gain access to accounts. However, with the availability of compromised credentials, this is no longer necessary.
- **Evading Detection**: SDEs are known for their ability to operate covertly for extended periods, characterized by patience in achieving their objectives, which can be complex and require stealthy execution over time. To avoid detection, SDEs employ various techniques, such as operating slowly and deliberately to prevent triggering alerts or arousing suspicion during regular monitoring. The research by Chung et al. employs advanced techniques such as Long Short-Term Memory (LSTM) and Deep Q-Learning (DQN) to dynamically predict and adapt to network traffic changes, emphasizing the growing risks associated with machine learning technologies. These technologies can autonomously initiate attacks and mimic user behavior. The study reveals that using machine learning for traffic monitoring, forecasting, and data exfiltration enables smarter, more covert operations. This represents a significant threat as malicious actors could leverage these capabilities to create data-driven malware and devise strategies to conceal their malicious activities and simulate accidental failures[17]. Specific capabilities designed to evade detection and minimize digital evidence include utilizing covert timing channels, exploiting zero-day vulnerabilities, and continuously evolving are key strategies for evading detection and thwarting forensic activities. Malware can be revised, upgraded, and reconnected if detected or performing sub optimally. Additionally, malicious traffic can be obfuscated through the injection of normal traffic or the creation of 'noise', while encryption and reverse-connect remote administration can also be employed. Hiding data, commands, and communications using disabled services and protocols such as DNS and HTTP tunnelling can further blend common commands and techniques.

- **Reconnaissance**: Current networks contain vast amounts of data and metadata. During the reconnaissance phase, malware can map networks, conduct OS fingerprinting, identify vulnerable systems, determine the common ports and services running, and locate digital assets. Attackers have access to numerous tools and techniques. Two common techniques include network sniffing through interfaces set to promiscuous mode and network scanning with tools like Nmap and Zenmap. Attackers must be careful to avoid detection, as network exploration generates significant and recognizable activity that may trigger alerts. New research showsthe possibility of transcoding hidden information in telephony protocols such as VOIP[18][19].
- **Communication with CnC**: The CnC center is the central hub of SDE operations. Once installed, malware on the infected machines communicates with and receives its direction from the CnC. It is a network of interconnected systems operated by malicious organizations. To facilitate persistence, it is designed for resiliency with sophisticated fail-over capabilities. These malicious networks are increasingly becoming cloud services for enhanced scalability[20], [21].

Communication between the malware and CnC center typically employs remote administration tools configured to evade detection[22]. Poison Ivy, a popular tool, operates in reverse-connect mode, which means there is no inbound traffic from CnC[23]. Regardless of the technology, the malware on the victim's network initiates all communications, essentially connecting to CnC to retrieve the next series of commands. Communication typically uses common protocols such as SMTP, HTTP, and DNS[24] and avoids detection using encryption [25], tunnelling[26], and timing techniques. With detailed knowledge of networks, systems, and file structures, it is now time to steal digital assets.

## 2.3. Phase 3 - Data Exfiltration

As the name suggests, the primary objective of SDE is to exfiltrate the data of value. Targeted assets often include login credentials, encryption keys, financial account information, email repositories, documents, and intellectual property. In addition, during the normal course of operations, malware may transfer metadata, such as file structures and network configuration data, to CnC for off-site analysis[17], [27].Interestingly, in less common role reversal, some malware is known to plant false or damaging information on the victimized network. They exfiltrate data unlawfully for offsite storage before encrypting it. Data exfiltration has become a standard procedure in almost all cases of ransomware attacks. In our work, we take up this currently most dangerous threat[28]. The motives for SDE vary, but a few reasons described in Table 1 for obtaining digital assets include the following:

Table 1: Example reasons behind data exfiltration attempts

| Why exfiltrate data? |
| --- |
| To sell the information |
| To sell successful attack methods |
| To publicly embarrass or humiliate |
| For competitive advantage |
| Ransom |

After identifying sensitive information, attackers typically conceal it using encryption, compression, file splitting, password protection, and various tunneling methods. In the RSA breach, the attackers collected the data into password-protected encrypted archive files and transferred them to an external staging server before picking up the data from the staging server and deleting it. Another technique for concealing data exfiltration is to create covert channels by embedding data or commands in common protocols such as HTTP, SMTP, and DNS. In the case of DNS tunneling, data is encoded into lower-level domains or less common DNS record types. These attacks can remain undetected for extended periods, but in this case, they have been successful. The target has been compromised, and the attackers have secretly obtained sensitive data. This type of malware is highly sophisticated, and many organizations lack the necessary resources to defend against such threats.

## 3. RECENT EXPLOITS

A prevalent view among information security professionals today is that most organizations are likely to be already compromised by SDE attacks without their awareness. Historically, these types of attacks have been utilized primarily in cyber warfare, targeting sectors such as defense, government, and the military. However, recent trends indicate that these sophisticated attacks have significantly expanded their targets, posing a threat to a broader range of industries [29], [30].

The expansion of SDE methods beyond traditional high-security domains into more public sectors underscores the need for advanced detection mechanisms and broader awareness of the threat landscape [31]. As these attacks become more common across various industries such healthcare, manufacturing, and finance. The need for a robust cybersecurity defenses and proactive threat detection has become increasingly critical [32].

Each exploit is highly targeted and, by nature, unique in motives, attacks, and operations. Understandably, public information about these attacks is not widely available. However, a brief review of recent high-profile attacks will help illustrate the breadth, depth, and organization of these attacks:

- **Microsoft 365 Phishing Scam**: In a cleverly orchestrated attack, cybercriminals sent emails containing a disguised HTML file in an Excel spreadsheet. When opened, this led the recipient to a fraudulent website designed to harvest Microsoft 365 login credentials under the pretense of a logged-out session. This incident highlights the reliance on human error and weak defenses, which have become more pronounced during the pandemic era [33].
- **Singapore Bank Phishing Attacks**: The Overseas Chinese Banking Corporation (OCBC) suffered a series of phishing attacks, which led to substantial financial losses. Attackers duped customers into surrendering their account details by phishing emails and swiftly transferring out funds to mule accounts [33].
- **Octo Tempest's Extortion Campaigns**: Leveraging sophisticated social engineering techniques; Octo Tempest targeted organizations across various sectors, including telecommunications and technology. By impersonating technical support or new employees, they gained unauthorized access and later extorted companies for the data stolen during these breaches [34].
- **TrickBot malware exploits**: Owing to its versatility in cyberattacks, TrickBot has spread primarily through spear-phishing campaigns, leading to extensive data exfiltration. It uses various methods, such as masquerading as legitimate software and injecting malicious code into system processes to evade detection and harvest sensitive information [35].
- In2023, APT carried out against government entities in the Asia-Pacific region compromised a particular type of USB drive that was used by these organizations to securely store and transmit data [36].

- BlindEagle aimed to attack both government organizations and individuals residing in South America. Although the primary objective of this threat actor is to gather intelligence, it also exhibited a desire to obtain financial information[36].
- The SolarWinds cyberattack, widely regarded as a major supply chain attack, has been traced back to APT29 (also known as Cozy Bear), a notorious Russian-sponsored APT group [37].
- Microsoft detected a state-sponsored APT group named Hafnium, which exploited vulnerabilities in the Microsoft Exchange Server to infiltrate email accounts and extract sensitive data. Microsoft discovered the group to be sponsored by the Chinese government [37].

These examples highlight the sophistication and prevalence of modern cyber threats and emphasize the urgent need for robust cybersecurity measures to protect against phishing and social engineering attacks. SDE poses a serious threat to a company's brand, reputation, and intellectual property. The next sections delve into the details of SDE, providing essential insights that will contribute to proposing a new paradigm for securing organizations against these sophisticated attacks.

## 4. RELATED WORK - EXISTING SECURITY MODELS

Much of the existing research, like that of Thompson et al.[38],is outdated, or not directly comparable to the MESA 2.0 security model, as in the case of Huang and Zhu[4]. For instance, researchers in various studies have employed multistage attack models, such as the cyber kill chain and MITRE ATT&CK, to detect APT and SDE from different perspectives. These play an important role but are limited to low level attack pattern detection and mitigation. MESA 2.0 provides holistic recommendations at an enterprise scale which could be implemented by an organization of any scale. Moreover, traditional literature reviews often face limitations, including a lack of comprehensiveness in addressing all challenges. Furthermore, previous reviews have not thoroughly evaluated commercially available technologies for APT detection. In contrast, the MESA 2.0 model offers a comprehensive framework that addresses the myriad challenges posed by cybersecurity threats to organizations. Unlike models that focus on specific types of attacks, MESA 2.0 acknowledges that attack patterns are constantly evolving. Within organizations, complex structures and disparate networks often lead to delayed and inconsistent responses to security threats, increasing vulnerability. Additionally, many security infrastructures evolve haphazardly over time, accruing numerous vulnerabilities that struggle to defend against evolving threats. A prevalent reactive mindset within organizations further hinders the development of effective security measures. By integrating behavioral insights and focusing on human factors, MESA 2.0 encourages organizations to adopt a more proactive and robust approach to cybersecurity.

MESA 2.0 advocates a broader approach to cybersecurity, addressing various aspects of threat detection and response. The model promotes continuous learning, tuning, and design updates to keep pace with the rapidly evolving cyber threat landscape, a strategy that is not as emphasized in other security models such as Sakr and El-Afifi[39] which presents a novel framework for confidential document leakage detection and prevention but limited to utilization of traditional forensic tools for data leakage prevention. MESA 2.0 model incorporates a comprehensive framework that not only focuses on evolving attack patterns but also emphasizes proactive measures, aspects that are not explicitly highlighted in other papers.

In summary, combating this new breed of malware requires an integrated response that addresses not only technological defenses but also organizational restructuring and a cultural shift towards proactive security practices. This comprehensive approach can better safeguard against sophisticated threats targeting the complex interdependencies of modern enterprise systems.

## 5. MESA 2.0 SECURITY MODEL

Traditional security measures are no longer adequate due to the advanced and evolving nature of threats. The new security approach emphasizes accepting the possibility of breaches and focusing on minimizing their impact through early detection and strong response strategies. This shift is driven by the recognition that APTs use sophisticated methods to bypass conventional security measures, often exploiting zero-day vulnerabilities, social engineering, and complex attack vectors.

Recent studies have advocated a combination of methodologies and technologies to better detect and respond to APTs. Techniques such as using open-source intelligence (OSINT) to gather data on potential threats and employing frameworks such as MITRE ATT&CK and the cyber kill chain to understand attack behaviors and tactics are crucial. These frameworks help in correlating the behavior of an APT across its attack lifecycle, thereby aiding in early detection and timely response to incidents[40], [41], [42]. Innovations in real-time attack detection frameworks, such as StreamSpot [43]and UNICORN[44], which do not require prior attack knowledge and exhibit low false-positive rates, represent significant advancements in defending against fileless and covert APT behaviors. Provenance graph-based approaches such as ProvDetector [45]have also been developed to trace hidden malicious activities within networks, further fortifying defenses against these sophisticated threats.Incorporating these advanced methodologies into existing security models does not imply abandoning risk-based prevention. Rather, it complements and strengthens them, ensuring a more resilient and adaptive security posture capable of handling the dynamic threat landscape of the APTs.

To adequately defend against advanced digital threats, a more comprehensive and integrated security model is needed that assumes a breach, rather than the prevailing security architecture design philosophy that focuses on prevention alone. Traditional security measures typically aim to block attackers before they can infiltrate systems or access sensitive data, but this method is becoming less effective as it is nearly impossible to stop all attacks. A new approach is necessary that incorporates a proactive mindset and leverages advanced technologies to detect and respond to breaches quickly. The MESA Security model, named after its contributors Muhammad, Eugene, Sanjeev, and Ankush, during a capstone project in 2012 at Northeastern University[46], [47] addresses this issue by advocating for a balanced focus on all critical aspects of enterprise security: information classification, prevention, detection, response, and continuous learning. At that time, this holistic approach ensured a more resilient and proactive security posture. More importantly, today this model requires significant updates to align with technologies that have evolved and are currently in use, such as cloud computing, IoT, and AI, which introduce new vulnerabilities. For instance, the proliferation of IoT devices significantly increases the attack surface, often with devices that lack security. This evolving landscape necessitates updates to security models, as reflected in Figure 2, which can adapt to and anticipate technological advancements and emerging threats [40].

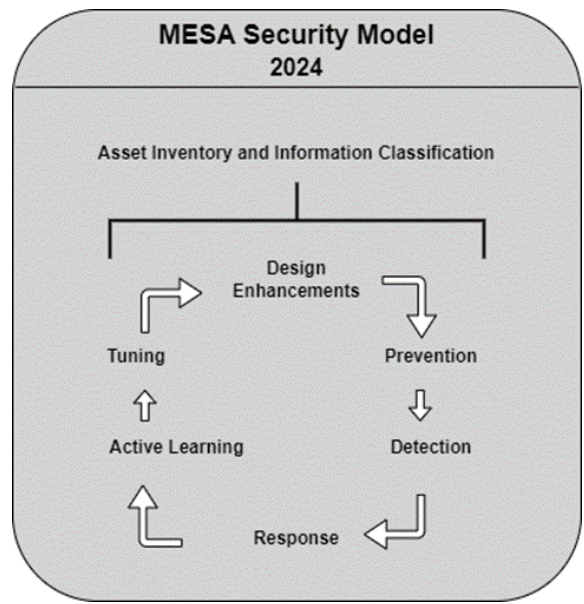

Figure 2 – MESA 2.0 Security Model

## 5.1. Asset Inventory and Information Classification

In today's cybersecurity landscape, the MESA 2.0 security model highlights the importance of a meticulous approach to information classification. This process starts with a crucial step: classifying data based on sensitivity and value. The MESA 2.0 security model employs a practical four-tier classification system for enterprise: Confidential, Private, Sensitive, and Public. The method prioritizes data by classification, ensuring the structured and resource-aware implementation of security measures. Moreover, having a comprehensive asset inventory and effective information classification is crucial for organizations to enhance their cybersecurity posture. By understanding all network assets and their criticality, organizations can identify vulnerabilities, implement targeted security measures, and proactively detect any anomalous activities. This approach enables early detection of stealth data exfiltration and APTs, helping prevent or minimize the impact of cyberattacks. Through automated techniques like event correlation, passive scanning, and network activity analysis, organizations can continuously monitor their assets, classify them accurately, and swiftly respond to potential threats. Such proactive measures are essential in today's dynamic and complex cyber threat landscape to safeguard sensitive data and critical infrastructure.

Kotenko et al. propose a method that aims to automatically track all digital and physical components in a computer network to ensure cybersecurity. This method employs a specialized technique to link and analyze system activities, which allows for efficient identification and management of these components. Furthermore, it utilizes a mathematical approach to categorize different characteristics of these components based on their *behavior or properties*, enhancing the organization and security of the network. The method also includes an *automated asset inventory based on event correlation*, which helps to identify network vulnerabilities and enables timely security measures and early detection of anomalies to prevent data exfiltration and APTs[48].

## 5.2. Prevention

The updated security model highlights the need for ongoing adaptation to stay ahead of the constantly evolving cyber threats. Implementing Information Security Management Systems

31

(ISMS) is crucial in protecting sensitive information and minimizing the risk of unauthorized access. Businesses must ensure accuracy and limit modifications only to authorized users. To meet regulatory requirements, it is essential to evaluate potential risks and develop a comprehensive ISMS that includes tailored policies and procedures. Common frameworks for an ISMS include the ISO/IEC 27001 standard, which provides standardized requirements for managing information security. Followed by comprehensive training for all staff to recognize and defend against spear phishing attacks, which target humans as the weakest link. High-risk or high-value individuals require specialized training and monitoring. Incident response teams play a crucial role in enhancing these training efforts by staying informed through industry resources, swiftly communicating threats, and ensuring timely actions. The team's role expands from reactive to proactive, maintaining vigilance against zero-day attacks. From a technical standpoint, traditional prevention methods are necessary, along with additional advanced defenses to counteract SDEs specifically. Such as leveraging data loss prevention (DLP) systems relying on text classification to assess the sensitivity level of data into categories mentioned in above section i.e. confidential, restricted, or public, using techniques ranging from binary classifications to more granular ones as referenced in Lal et al [49].

**APTs and multistage attack mitigation**

- The research by Che Mat et al. emphasizes the need to combine multi-stage attack-related behavior with periodic vulnerability assessments to prevent APT attacks by detecting and rectifying network deficiencies. It suggests using vulnerability scores and probability measures to improve the precision of pinpointing potential targets and prioritizing security measures. Furthermore, advanced visualization techniques are vital for charting the progression of APT attacks, allowing for early discovery and proactive defense techniques to safeguard the network from additional breaches [40].
- Provenance graphs: Provenance graph-based kernel-level auditing provides enhanced visibility and traceability within complex network environments, offering a strategic advantage in monitoring and securing network interactions[41].
- Continuous Security Training and Awareness: Ongoing training for all employees is essential to equip them with the knowledge to recognize potential threats and respond appropriately. This should cover the recognition of phishing attempts, the safe handling of data, and the importance of maintaining cybersecurity hygiene.
- Incident Response and Management: Having a proactive incident response team that is well versed in the latest APT tactics and equipped with up-to-date tools and procedures is vital. This team should be prepared to respond swiftly to potential threats, manage any incidents that occur, and ensure that the organization can recover quickly from attacks.
- Regular Security Assessments and Audits: Continuous assessments and audits of the security infrastructure help identify vulnerabilities that can be exploited by APTs. This includes regular penetration testing and vulnerability scans to ensure that all systems are secure and that any potential security gaps are addressed promptly.

**SDE Prevention**

- **Identifying Data Sources**: Recognizing all sources of data within the organization is the foundational step in safeguarding sensitive information and deploying cutting edge defense such as Extended Detection and Response (XDR) allowing for seamless monitoring of all data sources.
- **Determine Data Flows**: Understanding how data move and are processed within the organization helps in crafting effective protective measures.

- **Assign Data Ownership**: Designate individuals responsible for managing and securing specific datasets.
- **Apply Protection Measures**: Implement robust security protocols such as encryption and advanced access controls models such as Role Based Access Control (RBAC) and Attribute Based Access Control (ABAC) to protect data from unauthorized access and exfiltration.
- **Review Access Controls**: Periodically reviewing the access rights of individuals who have access to sensitive data and ensuring that they are current and in line with organizational policies, as well as implementing a comprehensive identity lifecycle management system that is well-developed.
- **Monitor Network Traffic**: Continuous monitoring and analysis of network traffic are critical for detecting and responding to unusual activities that may suggest data exfiltration attempts. Mundt and Baier have demonstrated the capability to counteract SDE attacks, which are often perpetrated by groups such as "SilverTerrier." They achieved this by integrating Cyber Threat Intelligence (CTI) and ISMS using the MITRE ATT&CK framework. The results showed that CTI effectively helps evaluate and prioritize cyber threats, thereby enhancing security measures against data theft. STIX (Structured Threat Information eXpression) and TAXII (Trusted Automated eXchange of Indicator Information) are utilized to format and share threat data efficiently. An example of an ISMS framework used in this integration is ISO/IEC 27001, which ensures systematic security measures to protect against data theft and exfiltration. These systems are continuously updated with CTI to improve security based on the latest intelligence. Additionally, the MITRE ATT&CK framework automates the integration of CTI and ISMS, providing a structured approach to understanding attack tactics and techniques that could lead to data exfiltration [50].
- **Zero-trust architecture and secure access service edge**: Research indicates that zero trust can effectively prevent or significantly mitigate APT attacks[51].

## 5.3. Detection

By integrating advanced technologies, continuous monitoring, and proactive strategies, organizations can enhance their detection capacity. MESA 2.0security model supports a balanced approach, enhancing robust detection alongside preventive measures, as relying solely on prevention may leave gaps in security.

Since the lateral movement of APTs greatly depends on the presence of vulnerabilities, it is crucial to consider this element to enhance the effectiveness of detection methods. To address this issue, Che Mat et al. suggests an improved APT detection technique that incorporates the relationship between APT attributes and network vulnerabilities. By integrating these two factors with attack path visualization, the approach improves the overall performance of APT detection mechanismsAPT focused threat hunting i.e. understanding and analyzing sequence of actions an APT might take, security systems can be tailored to detect unusual patterns more accurately[40].

Building on this foundational strategy, Willems et al. have proposed a novel approach that utilizes metadata from multiple network sessions to detect SDE rather than examining content directly. This metadata includes packet counts, session durations, and the intervals between sessions. By employing autoencoder neural networks, their model can identify anomalies and generate alerts based on predefined detection thresholds. The Network Exfiltration Detection System (NEDS) exemplifies efficiency in this context, analyzing network metadata to offer privacy while handling large data volumes. It trains autoencoders on typical traffic patterns using unsupervised learning to spot deviations. Users can fine-tune the system by adjusting the detection threshold and managing false positive rates, enhancing its capability to detect covert exfiltration attempts, including sophisticated tactics like DNS tunneling, without overwhelming false alarms[29].

Parallel to these developments, the MESA 2.0 security model pinpoints irregularities associated with malicious activities. It is crucial to investigate any atypical occurrences diligently. Recent advancements in AI and machine learning have significantly bolstered our ability to sift through vast amounts of network data to uncover anomalies that signal malicious activities. This technological leverage enables effective monitoring of potential attack scenarios, a method extensively detailed in research by Chung et al. In an enterprise environment, it's critical to strategically position detection systems near high-value targets or users who are most likely to be attacked. However, challenges arise with network-level monitoring—typically implemented at border routers—where it becomes difficult to distinguish between legitimate traffic and potential data exfiltration. Thus, deploying AI/ML solutions such as XDR directly on or near endpoints, especially those utilized by privileged users, is imperative. This tactic is particularly adept at identifying advanced cyber threats designed to bypass traditional detection frameworks[17].

Complementing these detection strategies is the crucial role of threat intelligence. It encompasses systematic research on potential adversaries, including the disclosure of their targets and tactics, and the discovery of emerging zero-day exploits and malicious domains. This intelligence not only fortifies an organization's defensive posture but also serves the broader data protection community through the dissemination of shared insights. Threat intelligence platforms play a pivotal role here, aggregating and analyzing data from multiple sources to equip organizations with up-to-date, actionable information. This helps them stay ahead in securing their networks against the latest attack vectors and the continually evolving tactics employed by cybercriminals.

## 5.4. Response

In MESA 2.0security model, the goal during the response phase is to minimize damage as efficiently as possible. Given the high stakes of SDEs, which target valuable digital assets, detectingAPT activities early in their lifecycle is crucial, allowing for preemptive actions before the threat escalates. A proper strategy and deployed solution should identify deviations from baseline behaviors and invoke an effective response to mitigate the impact of APTs and restore a secure and stable state. This involves several steps that depend on clear communication channels for proper escalation, with defined ownership and accountability. These steps are guided by a Security Incident Response Plan (SIRP) that includes established protocols for managing such incidents:

**Triage:** Advanced detection and efficient labeling of alerts enables rapid analysis and assessment of the situation.

**Containment and Eradication:** After identifying the extent of the problem, organizations are required to act swiftly to mitigate the threat and prevent any additional harm. This may involve automated resetting of exposed passwords, applying patches or other remedies to address security vulnerabilities, blocking malicious traffic and domains, and isolating infected systems. Following the containment and removal of the threat, a comprehensive forensic analysis is necessary to determine the root cause of the breach, assess the extent of the compromise, and gather evidence for legal and regulatory reasons. This usually entails analyzing logs, memory dumps, and network traffic to pinpoint indicators of compromise (IOCs) and reconstruct the timeline of the attack.

**Communication and Reporting:** While responding to an incident, it is essential to maintain clear communication with internal stakeholders, external partners, and regulatory authorities. Effective communication is crucial for building trust, ensuring compliance with legal and regulatory requirements, and managing the incident. By implementing robust incident response strategies, organizations can promptly respond to APT activities, thus minimizing potential impacts and safeguarding critical assets from threats.

The MESA 2.0 security model's main goal is to quickly halt the SDE activities. A dynamic incident plan should outline roles, responsibilities, and escalation procedures while equipping teams with tools for proactive preparation and efficient execution. Recognizing that SDEs are multi-vector attacks, incident teams must continuously adapt their defensive strategies by integrating both proactive and reactive measures. Thorough documentation of incidents via forensic analysis is crucial for long-term protection. It is essential for security programs to continuously learn from incidents to enhance their ability to respond to and anticipate future threats effectively.

## 5.5. Active Learning

Organizations must adapt and respond swiftly to emerging threats by incorporating active learning across all aspects of the MESA 2.0 security model, thus making it a crucial competency. Active learning involves the collection, analysis, sharing, and application of information from internal and external sources. This includes baselines, information classification, event correlation, threat intelligence networks, such as OSINT, Retail & Hospitality Information Sharing and Analysis Center (RH-ISAC) and Health Sector Information Sharing and Analysis Center (HS-ISAC), and forensic analysis. The transformation of this collected data into actionable intelligence is a powerful tool against SDE. One of the factors that can give adversaries an advantage is their utilization of sophisticated information-sharing networks. It is essential for organizations to share threat intelligence. An effective intelligence-sharing model promotes open contribution and learning from others. Although current threat intelligence networks are inadequate due to a lack of incentives, organizations should aim to participate and improve models for sharing threat intelligence. Earlier we discussed research by Mundt and Baier, which clearly reflected how organizations can enhance their security operations with up-to-date intelligence and responsive security management practices through dynamic interactions with other systems such as ISMS, effectively reducing the risk of data theft and ensuring compliance with international security standards. This approach not only streamlines the detection of data exfiltration attempts but also facilitates a proactive stance against cyber threats[50].

## 5.6. Tuning

Organizations must consistently update and adapt their security controls, policies, and procedures. This continuous refinement process not only addresses APTs and SDE-related threats, but also strengthens the overall resilience of security systems against newly discovered vulnerabilities and attacker tactics. Previously discussed autoencoder-based anomaly detection system required optimization to achieve high performance by adjusting configurable detection thresholds. To ensure a balance between false positive rates, this system must be regularly tuned, which enables it to adapt to various security requirements and improve its ability to detect SDE that typically occur slowly to avoid detection without triggering excessive false alarms for administrators[29].

The proactive stance is facilitated by regular updates to security policies such as *Forensic and Incident Response Preparedness*. In the event of a breach, a well-prepared incident response plan, regularly updated to address new types of security incidents, can significantly mitigate damage. This includes forensic tools and expertise in place to quickly analyze the nature of the attack, contain it, and recover from it. Mundt and Baier recommend "Proactive Simulations"[50]. The use of proactive simulations to test the effectiveness of implemented security measures against known attack techniques leveraging MITRE ATT&CK framework. These simulations help identify security gaps and reinforce defenses before actual attacks occur similar to table-top exercises.

Real-world incidents underscore the importance of this dynamic approach in policy management. For instance, the response to the NotPetya malware attack on various global entities highlights the need for robust backup and recovery policies as part of organizational security procedures. Entities with regularly updated policies that include advanced ransomware scenarios are better equipped to recover from the attack.

### 5.7. Design Enhancement

In the dynamic field of cybersecurity, regular updates to security architectures are vital for protecting against new and emerging threats. As cyber threats continue to evolve, insights from recent breaches must be quickly integrated into security frameworks to protect against future threats. These updates are crucial in various areas such as applications, networks, and access control systems.

- **Applications**: With the increasing complexity of software, the number of vulnerabilities has also increased. Regular enhancements in security design help to address newly identified exploits and weaknesses in applications. At a basic level this can involve patching known vulnerabilities, updating cryptographic standards, or refining access controls within the application itself. Mundt and Baier emphasize the importance of analyzing unintentional data leakage risks. They focus on data flows, data stores, and processes prone to information disclosure, especially in cases of data exfiltration. The Microsoft Threat Modeling Tool is useful in this analysis as it identifies all data flows across trust zone transitions, requiring mitigation actions to be identified and discussed. The tool provides an initial overview of a software's architecture, which can be further examined during architecture review board meetings. It is one part of a larger security ecosystem that includes Static Code Analysis and Security Testing Tools. Regular penetration testing is considered essential for all significant software releases[28].
- **Networks**: Network environments are constantly exposed to new forms of attacks, such as ransomware and APTs. Improving network security could involve deploying advanced IDS, enhancing firewall configurations, and adopting comprehensive network segmentation strategies such as micro-segmentation to limit lateral movement by malicious actors. Develop methods that enhance data protection strategies and tackle challenges related to scale and coverage. Advanced analytical approaches are necessary to provide deeper insights into data flows and identify potential exfiltration activities. There is a need for a system that continuously monitors sensitive data, revealing its content, movement context, and handler identities, especially as enterprises increasingly use multi-cloud models. Additionally, a dynamic data classification system is crucial for augmenting static/manual methods to better monitor and control sensitive data in transit. Many current solutions include predefined content patterns for personal identifiable information (PII), payment card industry (PCI) data, and protected health information (PHI), along with custom regex and keyword support.
  According to CrowdStrike's case study, a solution that differentiates data movements between managed and unmanaged applications can reduce false positives and improve SDE detection accuracy. This was evidenced by its implementation at a large enterprise,

  which alleviated user productivity and experience issues stemming from the complexity of traditional DLP systems [52].
- **Access**: The idea of "identity as the new perimeter" emphasizes the importance of robust access controls. Enhancements in this area often include implementing Multi-Factor Authentication (MFA), using more stringent identity verification methods, and updating access policies to adhere to the principle of least privilege, ensuring that users have only the necessary access to perform their roles. Judicious use of ABAC and RBAC models appropriately implemented should further reduce the risk of credentials misuse.

The incorporation of improved technology delivery mechanisms, including DevOps, Site Reliability Engineering (SRE), and automation, plays a crucial role in making these security design changes feasible and effective. DevOps practices promote the seamless integration of security into the software development lifecycle, advocating a "shift left" approach that addresses security earlier in the development process. SRE principles ensure that system reliability and security are considered together, thereby facilitating a more resilient infrastructure. Finally, automation significantly enhances security by enabling the rapid deployment of patches, automated security testing, and real-time threat detection and response, reducing human error, and increasing the speed of security responses. Therefore, the iterative improvement of security designs, supported by advanced methodologies and technologies, is not just necessary; it is fundamental to maintain robust defense mechanisms in a constantly changing threat landscape.

## 6. CONCLUSION

In conclusion, the swiftly changing landscape of cyber threats, characterized by advanced persistent threats (APTs) and stealth data exfiltration (SDE), calls for a substantial revamp of conventional enterprise security models. The research discussed in this paper highlights the inadequacy of traditional perimeter-based, prevention-focused defenses in the face of contemporary, complex cyber-attacks.

The transition towards a proactive security approach, as embodied by the MESA 2.0 security model, emphasizes the necessity of adopting a holistic strategy that integrates prevention, detection, response, continuous learning, tuning and security design updates. This model is designed to bolster an organization's capacity to proactively detect and respond to threats, thereby mitigating the repercussions of security breaches from SDE. Moreover, the importance of continuous security training and awareness, in conjunction with robust incident response plans, is pivotal in minimizing the impact of security breaches. The need for ongoing adaptation and fine-tuning of security controls is paramount, as demonstrated by the requirement for frequent updates to security architectures to tackle emerging threats.

Recent cyber incidents provide compelling evidence of the pressing need for enterprises to transition from a defensive stance that solely concentrates on prevention, to a more dynamic and proactive approach that includes preparation for inevitable breaches. This shift is vital for maintaining resilience against the increasingly sophisticated tactics employed by cyber adversaries. Furthermore, the incorporation of advanced technologies, such as Machine Learning (ML) and Artificial Intelligence (AI), into security systems, as proposed by the MESA 2.0 security model, signifies a progressive approach to cybersecurity. This ensures that security measures evolve concurrently with new technologies and threats, offering a comprehensive defense strategy capable of adapting to future challenges.

In summary, this study offers valuable insights and a strategic framework that could steer future security practices and policies. It ensures that enterprises are better prepared to confront and mitigate the effects of advanced persistent threats in the digital age. This comprehensive approach to cybersecurity not only addresses current threats but also anticipates future challenges, thereby providing a robust and adaptable defense mechanism. This forward-thinking approach is crucial in today's rapidly evolving digital landscape.

International Journal of Network Security & Its Applications (IJNSA) Vol.16, No.3, May 2024[37]  Hackerone, "Advanced Persistent Threat: Attack Stages, Examples & Mitigation." Accessed: Apr. 24, 2024. [Online]. Available: https://www.hackerone.com/knowledge-center/advanced-persistent-threats-attack-stages-examples-and-mitigation

[38]  B. Thompson, J. R. Morris-King, and H. Çam, "Controlling risk of data exfiltration in cyber networks due to stealthy propagating malware," *MILCOM 2016 - 2016 IEEE Mil. Commun. Conf.*, pp. 479–484, 2016, doi: 10.1109/MILCOM.2016.7795373.

[39]  H. A. Sakr and M. I. El-Afifi, "A Framework for Confidential Document Leakage Detection and Prevention," *Nile J. Commun. Comput. Sci.*, vol. 0, Apr. 2024, doi: 10.21608/njccs.2024.243509.1023.

[40]  N. I. Che Mat, N. Jamil, Y. Yusoff, and M. L. Mat Kiah, "A systematic literature review on advanced persistent threat behaviors and its detection strategy," *J. Cybersecurity*, vol. 10, no. 1, p. tyad023, Jan. 2024, doi: 10.1093/cybsec/tyad023.

[41]  Y. Wang, H. Liu, Z. Li, Z. Su, and J. Li, "Combating Advanced Persistent Threats: Challenges and Solutions," *IEEE Netw.*, pp. 1–1, 2024, doi: 10.1109/MNET.2024.3389734.

[42]  Q. Zou, X. Sun, P. Liu, and A. Singhal, "An Approach for Detection of Advanced Persistent Threat Attacks," *Computer*, vol. 53, no. 12, pp. 92–96, Dec. 2020, doi: 10.1109/MC.2020.3021548.

[43]  E. A. Manzoor, S. Momeni, V. N. Venkatakrishnan, and L. Akoglu, "Fast Memory-efficient Anomaly Detection in Streaming Heterogeneous Graphs." arXiv, Feb. 22, 2016. doi: 10.48550/arXiv.1602.04844.

[44]  X. Han, T. Pasquier, A. Bates, J. Mickens, and M. Seltzer, "UNICORN: Runtime Provenance-Based Detector for Advanced Persistent Threats," in *Proceedings 2020 Network and Distributed System Security Symposium*, 2020. doi: 10.14722/ndss.2020.24046.

[45]  Q. Wang *et al.*, "You Are What You Do: 27th Annual Network and Distributed System Security Symposium, NDSS 2020," *27th Annu. Netw. Distrib. Syst. Secur. Symp. NDSS 2020*, 2020, doi: 10.14722/ndss.2020.24167.

[46]  H. N. Duc, "LOW TECH HACKING, CISSP, NETWORK SCANNING E-BOOK 04/12 - Hakin9 - IT Security Magazine." Accessed: Apr. 24, 2024. [Online]. Available: https://hakin9.org/download/low-tech-hacking-cissp-network-scanning-e-book-0412/

[47]  M. Higgins, S. Pratap Singh, E. Tyrrell, A. Patankar, and M. Taibah, "The Evolving Threat Landscape: Is it Time for a Paradigm Shift in the Enterprise Security Model?"

[48]  I. Kotenko, E. Doynikova, A. Fedorchenko, and V. Desnitsky, "Automation of Asset Inventory for Cyber Security: Investigation of Event Correlation-Based Technique," *Electronics*, vol. 11, no. 15, Art. no. 15, Jan. 2022, doi: 10.3390/electronics11152368.

[49]  A. Lal, A. Prasad, A. Kumar, and S. Kumar, "Data Exfiltration: Preventive and Detective Countermeasures." Rochester, NY, Feb. 10, 2022. doi: 10.2139/ssrn.4031852.

[50]  M. Mundt and H. Baier, "Towards Mitigation of Data Exfiltration Techniques Using the MITRE ATT&CK Framework," in *Digital Forensics and Cyber Crime*, P. Gladyshev, S. Goel, J. James, G. Markowsky, and D. Johnson, Eds., Cham: Springer International Publishing, 2022, pp. 139–158. doi: 10.1007/978-3-031-06365-7_9.

[51]  B. Karabacak and T. Whittaker, "Zero Trust and Advanced Persistent Threats: Who Will Win the War?," *Int. Conf. Cyber Warf. Secur.*, vol. 17, no. 1, pp. 92–101, Mar. 2022, doi: 10.34190/iccws.17.1.10.

[52]  CrowdStrike, "A Modern Approach to Stopping Data Exfiltration | CrowdStrike," crowdstrike.com. Accessed: May 04, 2024. [Online]. Available: https://www.crowdstrike.com/resources/white-papers/modern-approach-to-stopping-data-exfiltration/
40